\documentclass[12pt]{article}

\usepackage{latexsym}
\usepackage{amssymb}
\usepackage{amsmath}

\newtheorem{theorem}{Theorem}

\newcommand{\niklabel}[1]{\label{#1}}

\def\proof{\medbreak\noindent{\bf Proof}}

\newcommand{\cl}{{{\cal C}\!\ell}}

\def\{{\lbrace}
\def\}{\rbrace}

\def\st{\stackrel}
\def\for{\quad\hbox{for}\quad}

\def\ww{\wedge\ldots\wedge}

\def\R{{\mathbb R}}
\def\C{{\mathbb C}}
\def\F{{\mathbb F}}

\def\U{{\rm U}}
\def\SU{{\rm SU}}

\def\E{{\cal E}}
\def\Tr{{\rm Tr}}

\def\diag{{\rm diag}}
\def\even{{\rm even}}
\def\odd{{\rm odd}}
\def\det{{\rm det}}

\def\Sp{{\rm Sp}}
\def\sp{{\rm sp}}

\def\be{\begin{equation}}
\def\ee{\end{equation}}
\def\Det{{\rm Det}}

\def\Mat{{\rm Mat}}

\newcommand{\dimension}{{\rm dim\,}}

\def\be{\begin{equation}}
\def\ee{\end{equation}}

\def\SW{{\rm SW}}
\def\sw{{\rm sw}}
\def\W{{\rm W}}
\def\w{{\rm w}}

\begin{document}


\author{Nikolai Marchuk,  Roman Dyabirov}

\title{A symplectic subgroup of a pseudounitary group as a
subset of Clifford algebra}

\maketitle

\begin{abstract}
Let Cl1(1,3) and Cl2(1,3) be the subsets of elements of the
Clifford algebra Cl(1,3) of ranks 1 and 2 respectively. Recently
it was proved that the subset Cl2(p,q)+iCl1(p,q) of the complex
Clifford algebra can be considered as a Lie algebra. In this paper
we prove that for p=1, q=3 the Lie algebra Cl2(p,q)+iCl1(p,q) is
isomorphic to the well known matrix Lie algebra sp(4,R) of the
symplectic Lie group Sp(4,R). Also we define the so called
symplectic group of Clifford algebra and prove that this Lie group
is isomprphic to the symplectic matrix group Sp(4,R).
\end{abstract}

\section*{Introduction}
In \cite{MarShi} was considered the complex Clifford algebras
$\cl(p,q)$ and was found subsets of Clifford algebra closed with
respect to the commutator $[U,V]=UV-VU$. For all positive integer
$p,q$ ($p+q\geq2$) the subset $\cl_2^\R(p,q)\oplus i\cl_1^\R(p,q)$
of Clifford algebra $\cl(p,q)$ can be considered as a Lie algebra.
In this paper we prove that for $p=1,q=3$ the Lie algebra
$\cl_2^\R(p,q)\oplus i\cl_1^\R(p,q)$ is isomorphic to the well known
matrix Lie algebra $\sp(4,R)$ of the symplectic Lie group
$\Sp(4,R)$. Also we define the so called symplectic group of
Clifford algebra and prove that this Lie group is isomprphic to the
symplectic matrix group $\Sp(4,R)$.


\section{Clifford algebras $\cl(p,q)$}
\niklabel{section330}

Let F be the field of real numbers $\R$ or the field of complex
numbers $\C$ and let $n$ be a natural number. Consider $2^n$
dimensional vector space $\E$ over the field $\F$ with a basis

\begin{equation}
e,\,e^a,\,e^{a_1 a_2},\ldots, e^{1\ldots n}, \quad a_1<a_2<\cdots,
\niklabel{basis:cliff}
\end{equation}
with elements numbered dy ordered multi-indices of length from $0$
to $n$. Indices $a,a_1,\ldots$ take values from $1$ to $n$.

Let $p,q$ be nonnegative integer numbers and $p+q=n$. Consider a
diagonal matrix of dimension $n$

\begin{equation}
\eta=\eta(p,q)=\diag(1,\ldots,1,-1,\ldots,-1),
\niklabel{eta}
\end{equation}
with $p$ pieces of $1$ and $q$ pieces of $-1$ on the diagonal. By
$\eta^{ab}=\eta_{ab}$ denote elements of $\eta$.

Following rules define product $U,V\to UV$ of elements of the
vector space $\E$:
\begin{enumerate}
\item For any $U,V,W\in\E$
\begin{eqnarray*}
&&U(V+W)=U V+U W,\quad
(U+V) W=U W+V W,\\
&&(U V) W=U(V W);
\end{eqnarray*}
\item $e U=U e=U$, for any $U\in\E$;

\item $e^a  e^b+e^b e^a=2\eta^{ab}e$, for $a,b=1,\ldots,n$;
\item $e^{a_1}\ldots e^{a_k}=e^{a_1\ldots a_k}$, for $1\leq
a_1<\cdots<a_k\leq n$.
\end{enumerate}

The operation of multiplication converts vector space $\E$ to an
algebra. This algebra is called {\it the Clifford
algebra}\index{Clifford algebra} and denoted by $\cl^\F(p,q)$ (if
$\F=\C$, then $\cl(p,q)=\cl^\C(p,q)$). For $p=n, q=0$ we use
notation $\cl^\F(n)=\cl^\F(n,0)$. Elements $e^a$ are called {\it
generators}\index{generators} of Clifford algebra $\cl^\F(p,q)$.
$\cl^\R(p,q)$ is real Clifford algebra and $\cl(p,q)$ is complex
Clifford algebra.

Any element $U\in\cl^\F(p,q)$ can be written in the form
\begin{equation}
U=ue+u_a e^a+\sum_{a_1<a_2} u_{a_1 a_2}e^{a_1 a_2}+\ldots
+u_{1\ldots n}e^{1\ldots n}
\niklabel{U:decomp1}
\end{equation}
with coefficients $u,u_a,u_{a_1 a_2},\ldots,u_{1\ldots n}\in\F$,
which numbered by ordered multi-indices of length from $0$ to $n$.

Denote by $\cl^\F_k(p,q)$, $k=0,\ldots n$, subspaces of the vector
space $\cl^\F(p,q)$that span over basis elements $e^{a_1\ldots
a_k}$.  Elements of $\cl^\F_k(p,q)$ are called elements of rank
$k$. Sometimes it is suitable to denote
$\st{k}{U}\in\cl^\F_k(p,q)$. We have
$$
\cl^\F(p,q)=\cl^\F_0(p,q)\oplus\ldots\oplus \cl^\F_n(p,q)
=\cl^\F_\even(p,q)\oplus\cl^\F_\odd(p,q),
$$
where
\begin{eqnarray*}
\cl^\F_\even(p,q)&=&\cl^\F_0(p,q)\oplus\cl^\F_2(p,q)\oplus\ldots,\\
\cl^\F_\odd(p,q)&=&\cl^\F_1(p,q)\oplus\cl^\F_3(p,q)\oplus\ldots,
\end{eqnarray*}
and
$$
\dimension \cl^\F_k(p,q)=C^k_n,\quad\dimension\cl^\F_\even(p,q)=
\dimension\cl^\F_\odd(p,q)=2^{n-1},
$$
$C^k_n$ are binomial coefficients. Let us take the antisymmetric
coefficients $u_{a_1\ldots a_k}=u_{[a_1\ldots a_k]}$. Consider an
element
\begin{eqnarray*}
\st{k}{U}&=&\sum_{a_1<\cdots<a_k}u_{a_1\ldots a_k}e^{a_1\ldots
a_k}= \sum_{a_1<\cdots<a_k}u_{a_1\ldots a_k}e^{a_1}\ldots e^{a_k}.
\end{eqnarray*}
We have
$$
\st{k}{U}=\frac{1}{k!}u_{a_1\ldots a_k}e^{a_1}\ldots e^{a_k}=
\frac{1}{k!}u_{a_1\ldots a_k}e^{a_1}\ww e^{a_k}.
$$
for U from (3).


\section{A Hermitian conjugation operation}
Using the projection operation onto the one-dimensional subspace
$\cl_0(p,q)$, we get {\em a trace} of an element $U\in\cl(p,q)$
$$
\Tr\,:\,\cl(p,q)\to\C,
$$

where
$$
\Tr(ue+u_a e^a+\ldots)=u.
$$

This operation has the following properties:
$$
\Tr(UV)=\Tr(VU),\quad \Tr([U,V])=0.
$$

Now we introduce the structure of unitary (Euclidian) space on
Clifford algebra \cite{MarShi}. Let $\dagger :
\cl^\F(p,q)\to\cl^\F(p,q)$ be a linear Hermitian conjugation
operation of the Clifford algebra elements. Following rules define
a Hermitian conjugation operation of Clifford algebra elements:

$$
(e^{i_1\ldots i_k})^\dagger=e_{i_k}\ldots e_{i_1},\quad \hbox{and}
\quad \lambda^\dagger=\bar\lambda,\quad \forall\lambda\in\C,
$$
$$
 (U+V)^\dagger=V^\dagger + U^\dagger \quad
\forall U,V\in\cl^\F(p,q), \niklabel{ea:hermconj}\\
$$

where $e_a=\eta_{ab}e^b$. We say that
$\dagger:\cl^\F(p,q)\to\cl^\F(p,q)$ is the {\it operation of
Hermitian conjugation} of Clifford algebra elements. It is easy to
see that
$$
(U V)^\dagger=V^\dagger U^\dagger,\quad U^{\dagger\dagger}=U.
$$
Now we can define a Hermitian (Euclidian) scalar product of
Clifford algebra elements by the formula
$$
(U,V)=\Tr(V^\dagger U).
$$
In this case we have
$$
(e^{i_1\ldots i_k},e^{i_1\ldots i_k})= \Tr(e_{i_k}\ldots
e_{i_1}e^{i_1}\ldots e^{i_k})=\Tr(e)=1,
$$
(no summation w.r.t. $i_1,\ldots,i_k$). Basis (\ref{basis:cliff})
of $\cl^\F(p,q)$ is orthonormal with respect to this scalar
product and property (7) is valid
\begin{equation}
(U,U)=\sum_{k=0}^n \sum_{a_1<\ldots<a_k}|u_{a_1\ldots a_k}|^2>0.
\niklabel{scal:pos}
\end{equation}
For generators $e^a$ formula (6) gives
\begin{eqnarray}
(e^a)^\dagger&=&e^a\for a=1,\ldots,p;
\niklabel{ea:hermconj}\\
(e^a)^\dagger&=&-e^a\for a=p+1,\ldots,n.\nonumber
\end{eqnarray}


\section{\sloppy Pseudounitary groups of \\Clifford algebras}
\niklabel{section430}

Let $r,s$ be nonnegative integers, $r+s>0$. Consider a Lie group
of pseudounitary matrices and a Lie group of special pseudounitary
matrices:
\begin{eqnarray*}
\U(r,s)&=&\{U\in\Mat(n,\C): U^\dagger \beta U=\beta\},\\
\SU(r,s)&=&\{U\in\U(r,s): \det\,U=1\},
\end{eqnarray*}
where $n=r+s$ and $\beta=\diag(1,\ldots,1,-1,\ldots,-1)$ is a diagonal matrix with $r$ pieces of $1$ and
$s$ pieces of $-1$ on the diagonal.

Now let us consider the four sets of Clifford algebra elements:
\begin{eqnarray*}
\W(\cl(p,q))&=&\{U\in\cl(p,q): U^*U=e\},\\
\SW(\cl(p,q))&=&\{U\in\W(\cl(p,q)): \Det\,U=1\},\\
\w(\cl(p,q))&=&\{U\in\cl(p,q): U^*=-U\},\\
\sw(\cl(p,q))&=&\{U\in\w(\cl(p,q)): \Tr\,U=0\}.
\end{eqnarray*}
Note that the sets $\W(\cl(p,q))$ and $\SW(\cl(p,q))$ are closed
with respect to the Clifford product and the sets
 $\w(\cl(p,q))$ and $\sw(\cl(p,q))$
are closed with respect to the commutator. The sets $\W(\cl(p,q))$
and $\SW(\cl(p,q))$ can be considered as Lie groups and the sets
$\w(\cl(p,q))$ and $\sw(\cl(p,q))$ can be considered as Lie
algebras of the Lie groups $\W(\cl(p,q))$ and $\SW(\cl(p,q))$.

The group $\W(\cl(p,q))$ is called {\it the pseudounitary group of
Clifford algebra}\index{pseudounitary group of the Clifford
algebra} $\cl(p,q)$. The group $\SW(\cl(p,q))$ is called {\it the
special pseudounitary group of Clifford algebra}.
\medskip

 \section{Symplectic Lie groups and their Lie algebras} Let us consider a {\em real symplectic Lie group} of
$n \times n$-matrices, $n=2m$ and let us consider the real Lie
algebra of this group
\begin{eqnarray*}
\Sp(n,\R) &=& \{U\in\Mat(n,\R) : U^T J U=J\},\\
\sp(n,\R) &=& \{u\in\Mat(n,\R) : u^T J =-J u\},
\end{eqnarray*}
where $U^T$ is a transpose matrix, $J$ is the matrix
$$
J=\left( \begin{array}{cc} 0 & -I_m\\ I_m & 0\end{array}\right)
$$
and $I_m$ is an identity $m\times m$- matrix. Note that
$J^2=-{\bf1}$ (${\bf1}$ is the identity $4\times4$ matrix).
Mostafazadeh \cite{Most} shows that the symplectic group
$\sp(2n,R)$ is a subgroup of the pseudounitary group $\U(n,n)$.
\medskip

\section{A symplectic group of Clifford algebra} Now define the two sets of Clifford algebra
elements
\begin{eqnarray*}
\Sp(\cl(1,3)) &=& \{V\in\cl^\R_\even(1,3)\oplus i\cl^\R_\odd(1,3) : V^*V=e\},\\
\sp(\cl(1,3)) &=& \{v\in i\cl^\R_1(1,3)\oplus \cl^\R_2(1,3)\}.
\end{eqnarray*}
We have $\Sp(\cl(1,3))\subset\W(\cl(1,3))$. Suppose that this set
is closed with respect to the product; then $\Sp(\cl(1,3))$ can be
considered as a group (the Lie group) with respect to the product.
This group is called {\em the symplectic group of Clifford algebra
$\cl(1,3)$}\index{the symplectic group of Clifford algebra}. The
relation of this group with the matrix symplectic group will be
explained in Theorem \ref{sp:theorem}.

In \cite{MarShi} we see that the set
$\sp(\cl(1,3))\subset\w(\cl(1,3))$ is closed with respect to the
commutator $[u,v]=u v-v u$. So $\sp(\cl(1,3))$ is a Lie algebra
with respect to the commutator.

\begin{theorem}.\niklabel{sp:theorem} The group $\Sp(\cl(1,3))$ is isomorphic to the group $\Sp(4,\R)$ and the Lie algebra $\sp(\cl(1,3))$ is isomorphic to the Lie algebra $\sp(4,\R)$:
\begin{eqnarray}
\Sp(\cl(1,3)) &\simeq& \Sp(4,\R),\niklabel{sp:sp}\\
\sp(\cl(1,3)) &\simeq& \sp(4,\R).\nonumber
\end{eqnarray}
\end{theorem}

\proof. It is well known \cite{Lounesto} that the complex Clifford
algebra $\cl(1,3)$ is isomorphic to the algebra of complex
matrices $\Mat(4,\C)$. Let $e^a$, $a=0,1,2,3$ be generators and
$e$ be the identity element of Clifford algebra $\cl(1,3)$.
Consider the matrix representation $\gamma$ of Clifford algebra
elements
$$
\gamma\ :\ \cl(1,3)\ \to\ \Mat(4,\C),
$$
such that
\begin{eqnarray*}
\gamma(A+B) &=& \gamma(A)+\gamma(B),\\
\gamma(\lambda A) &=&\lambda\gamma(A),\\
\gamma(A B) &=& \gamma(A)\gamma(B),\\
\gamma(e) &=& {\bf1},
\end{eqnarray*}
where $A,B$ are arbitrary elements of $\cl(1,3)$; $\lambda$ is a
complex number; ${\bf1}$ is the identity $4\times4$-matrix and
generators of Clifford algebra represented by the following
matrices:
\begin{eqnarray*}
\gamma(e^0) &=\left(\begin{array}{cccc}
0 & 0 & i & 0 \\
0 & 0 & 0 & i \\
-i & 0 & 0 & 0 \\
0 & -i & 0 & 0
\end{array}\right),\qquad
\gamma(e^1) &=\left(\begin{array}{cccc}
0 & -i & 0 & 0 \\
-i & 0 & 0 & 0 \\
0 & 0 & 0 & i \\
0 & 0 & i & 0
\end{array}\right),\\
\gamma(e^2) &=\left(\begin{array}{cccc}
0 & 0 & i & 0 \\
0 & 0 & 0 & i \\
i & 0 & 0 & 0 \\
0 & i & 0 & 0
\end{array}\right),\qquad
\gamma(e^3) &=\left(\begin{array}{cccc}
-i & 0 & 0 & 0 \\
0 & i & 0 & 0 \\
0 & 0 & i & 0 \\
0 & 0 & 0 & -i
\end{array}\right).
\end{eqnarray*}

We say that the matrix representation $\gamma$ is a {\em modified
Majorana's representation}\index{modified Majorana's
representation}. Let us remind that the representation $\gamma$
defines the isomorphism of  $\cl(1,3)$ and $\Mat(4,\C)$.

Generators $e^a$ of Clifford algebra satisfy the following
formulas
$$
e^a e^b+e^b e^a=2\eta^{ab}e,\quad a,b=0,1,2,3.
$$
So the matrices $\gamma^a=\gamma(e^a)$, $a=0,1,2,3$ satisfy the
conditions
$$
\gamma^a\gamma^b+\gamma^b\gamma^a=2\eta^{ab}{\bf1},\quad
a,b=0,1,2,3.
$$
Note that the sixteen matrices
$$
{\bf1},\ \gamma^a,\ \gamma^a\gamma^b,\ \gamma^a\gamma^b\gamma^c,\
\gamma^0\gamma^1\gamma^2\gamma^3,\quad a<b<\ldots
$$
are linearly independent and form the  basis of the matrix algebra
$\Mat(4,\C)$.

Note also that the matrices $\gamma^a$ satisfy to the conditions
$$
(\gamma^0)^\dagger=\gamma^0,\quad (\gamma^k)^\dagger=-\gamma^k,\quad
k=1,2,3.
$$
Therefore the representation $\gamma$ is consistent with the
Hermitian conjugation operation
$$
\gamma(A)^\dagger=\gamma(A^\dagger),
$$
where $\gamma(A)^\dagger$ is the Hermition conjugated matrix, and
$A^\dagger=e^0A^*e^0$ is the Hermitian conjugated element of
Clifford algebra $\cl(1,3)$.

Denote by $\gamma^{-1}$ the inverse map
$$
\gamma^{-1}\ :\ \Mat(4,\C) \to \cl(1,3).
$$
Matrices $i\gamma^a$ are real, therefore the maps $\gamma,
\gamma^{-1}$ give the isomorphism
$$
\cl^\R_\even(1,3)\oplus i\cl^\R_\odd(1,3)\simeq\Mat(4,\R).
\label{simeq}
$$

Our the representation $\gamma$ is such that
$$
 J=i\gamma(e^0),
$$
where the matrix $J$ is used in the definition of the group of
symplectic matrices.

For any $A\in\cl(1,3)$ we have
$$
A^\dagger=e^0A^*e^0, or
A^*=e^0A^\dagger e^0.
$$
Therefore for any element $A\in\cl^\R_\even(1,3)\oplus
i\cl^\R_\odd(1,3)$ we have
$$
\gamma(A^*)=\gamma(e^0)\gamma(A^\dagger)\gamma(e^0)= -J\gamma(A)^T
J.
$$
Consider the definition of group $\Sp(\cl(1,3))$. The condition
$A^*A=e$ leads to the relation
$$
\gamma(A^*)\gamma(A)=\gamma(e).
$$
Using this relation and conditions
$$ \gamma(A^*)=-J\gamma(A)^T J,\quad
J^2=-{\bf1}$$ we get
$$
\gamma(A)^T J\gamma(A)=J.
$$
From the definition of the group $\Sp(4,\R)$ it follows that the
group $\Sp(\cl(1,3))$ is isomorphic to the group $\Sp(4,\R)$
$$\Sp(\cl(1,3))\simeq\Sp(4,\R).$$
Let us remind the definition of Lie algebra $\w(\cl(1,3))$ of the
pseudounitary group $\W(\cl(1,3))$
$$
\w(\cl(1,3))=\{w\in\cl(1,3) : w^*=-w\}.
$$
This definition is equivalent to the following:
$$
\w(\cl(1,3))=\{w\in i\cl^\R_0(1,3)\oplus i\cl^\R_1(1,3)\oplus
\cl^\R_2(1,3)\oplus \cl^\R_3(1,3)\oplus i\cl^\R_4(1,3)\}.
$$
Note that the definition of Lie algebra $\sp(\cl(1,3))$ can be
written in the form
$$
\sp(\cl(1,3))=\{w\in\w(\cl(1,3)) : w\in\cl^\R_\even(1,3)\oplus
i\cl^\R_\odd(1,3)\}.
$$

Hence, $\sp(\cl(1,3))$ is the intersection of following sets:
\begin{eqnarray*}
&&\{i\cl^\R_0(1,3)\oplus i\cl^\R_1(1,3)\oplus \cl^\R_2(1,3)\oplus
\cl^\R_3(1,3)\oplus i\cl^\R_4(1,3)\}\\
&&\cap \{\cl^\R_\even(1,3)\oplus i\cl^\R_\odd(1,3)\}.
\end{eqnarray*}
Any element of the Lie algebra $w\in\cl^\R_\even(1,3)\oplus
i\cl^\R_\odd(1,3)$ represents as a real matrix. Therefore using
the matrix representation $\gamma$ for the expression $w^*=-w$ we
get the relations
$$
-J\gamma(w)^T J=-\gamma(w), \quad \hbox{or} \quad \gamma(w)^T
J=-J\gamma(w).
$$
It is clear that the last expression defines the Lie algebra
$\sp(4,\R)$. Hence the Lie algebra
 $\sp(\cl(1,3))$ is isomorphic to the Lie algebra $\sp(4,\R)$. This completes the proof.


\begin{thebibliography}{99}

\bibitem{MarShi} Marchuk N. G., Shirokov D. S. {\sl Unitary spaces on Clifford
algebras}, Adv. appl. Clifford alg., 18, (2008), pp.237-254,
 arXiv:0705.1641v1 [math-ph], (2007).

\bibitem{Hestenes} Hestenes D., {\sl Space-Time Algebra},
Gordon and Breach, New York, (1966).\par

\bibitem{Lounesto} Lounesto P., {\sl Clifford Algebras and Spinors},
Cambridge Univ. Press (1997, 2001)\par

\bibitem{Most} Mostafazadeh A., {\sl Pseudo-Unitary Operators and Pseudo-Unitary Quantum
Dynamics}, arXiv:0302050v2 [math-ph], (2003).
\end{thebibliography}
\end{document}